\documentclass[aps,prl,preprint,groupedaddress,amsfonts,showpacs]{revtex4}
\usepackage{graphicx,color}
\usepackage{bm}
\usepackage[latin1]{inputenc}

\begin{document}

\title{Quantum transport in ferromagnetic Permalloy nanostructures}

\author{D. Neumaier}
\email{daniel.neumaier@physik.uni-regensburg.de}
\author{A. Vogl}
\author{J. Eroms}
\author{D. Weiss}

\affiliation{Institut f\"{u}r Experimentelle und Angewandte Physik,
Universit\"{a}t Regensburg, 93040 Regensburg, Germany}

\date{\today}

\begin{abstract}
We studied phase coherent phenomena in mesoscopic Permalloy samples
by exploring low temperature transport. Both, differential
conductance as a function of bias voltage and magnetoconductance of
individual wires display conductance fluctuations. Analysis of these
fluctuations yields a phase coherence length of $\sim250$ nm at 25
mK as well as a $1/\sqrt{T}$ temperature dependence. To suppress
conductance fluctuations by ensemble averaging we investigated low
temperature transport in wire arrays and extended Permalloy films.
In these samples we have measured conductance corrections which stem
from electron-electron interaction (EEI) but attempts to detect
signatures of weak localization were without success.
\end{abstract}


\maketitle
\section{Introduction}

In nanoscale samples the conductance is affected by quantum
interference effects at sufficiently low temperatures, including
weak localization (WL) \cite{Bergmann}, universal conductance
fluctuations (UCF) \cite{Lee} or Aharonov-Bohm oscillations
\cite{Aharonov}. All these effects rely on the electron's wave
nature and require phase coherent transport over a certain distance.
In nonmagnetic metals interference effects have been investigated
intensely over the last two decades (for a review see, e.g.,
reference \cite{Imry}), showing that the electrons can propagate up
to several microns without losing their phaseinformation, although
their mean free path is much smaller. For ferromagnetic metals, in
contrast, only a few experimental works on phase coherent phenomena
exist (e.g. \cite{Wei,Kasai,Lee3}). While Aharonov-Bohm oscillations
or universal conductance fluctuations have been observed in
Permalloy \cite{Kasai} and cobalt devices \cite{Wei}, the
suppression of weak localization is still an open question. Up to
now no clear signature of weak localization was found in any
ferromagnetic metal. For example, the conductance of Co
\cite{Brands,Wei}, Co/Pt multilayers \cite{Brands3}, Fe
\cite{Brands2} and Ni \cite{Ono} was not affected by weak
localization. In these materials a decreasing conductance with
decreasing temperature was ascribed to electron-electron interaction
\cite{Brands,Wei,Brands3,Brands2,Ono}. However, in the ferromagnetic
semiconductor (Ga,Mn)As \cite{Ohno} weak localization corrections
could be observed quite recently \cite{WL,Rozkinson}. In conmtrast
to ferromagnetic metals the internal magnetic induction of (Ga,Mn)As
is rather small. In this work we will investigate universal
conductance fluctuations, weak localization and electron-electron
interaction in Permalloy nanostructures.

\section{Sample preparation and measurement technique}

For the experiments we fabricated single wires, arrays of wires
connected in parallel and thin film areas using electron beam
lithography (a Zeiss electron microscope equipped with a nanonic
pattern generator) and thermal evaporation of Permalloy. The
contacts to the samples were made out of gold after a brief in-situ
ion beam etching to remove the oxide. An electron micrograph of a
145 nm long, 35 nm wide and 15 nm thick Permalloy wire is shown in
figure 1a.

All measurements have been performed in a top-loading dilution
refrigerator using standard 4 probe lock-in techniques. The external
magnetic field was always applied perpendicular to the plane. To
avoid heating of the electrons careful shielding as well as small
measuring currents (dependending on the sample's resistance ranging
from 4 nA to 400 nA) were crucial. To measure the differential
resistance we also used standard lock-in techniques. A DC-voltage
$U_{DC}$ was superimposed upon a comparably small AC-voltage
$U_{AC}$ ($5$ $\mu$V) having a frequency of 417 Hz. The conductance
$G$ was obtained by inverting the resistance $R$ of the samples
$G=1/R$, while the differential conductance was obtain by inverting
the differential resistance $G^d=1/R^d=1/($d$U/$d$I)$.

\section{Universal conductance fluctuations in single wires}

Universal conductance fluctuations (UCF) result from correlations
between different transmission paths through a disordered mesoscopic
sample (\cite{Washburn} and references therein). Hence the
conductance $G$ of a mesoscopic sample is sensitive to the impurity
configuration. Changing the applied magnetic field $B$ or the
applied bias voltage $U$ leads to aperiodic conductance fluctuations
\cite{Washburn}. While changing in the magnetic field leads to a
relative phase shift of different paths due to the
Aharonov-Bohm-effect, a change in bias voltage changes the
electron's energy and thus the corresponding wavelength. For
observing UCF, phase coherence is necessary. If the sample is larger
than the phase coherence length in one or more spatial dimensions
the fluctuations get damped until the classical value of conductance
is reached \cite{Lee}. Hence sufficiently small samples and low
temperatures are required to observe universal conductance
fluctuations.

Here we investigate conductance fluctuations in differential
conductance $G^d(U)$ and in magnetoconductance $G(B)$ of two single
Py wires having a length of 145 nm and 330 nm respectively. The
width of both wires is 35 nm and the thickness is 15 nm. An electron
micrograph of the 145 nm long wire is shown in figure 1a.

\begin{figure}
\includegraphics[width=0.55\linewidth]{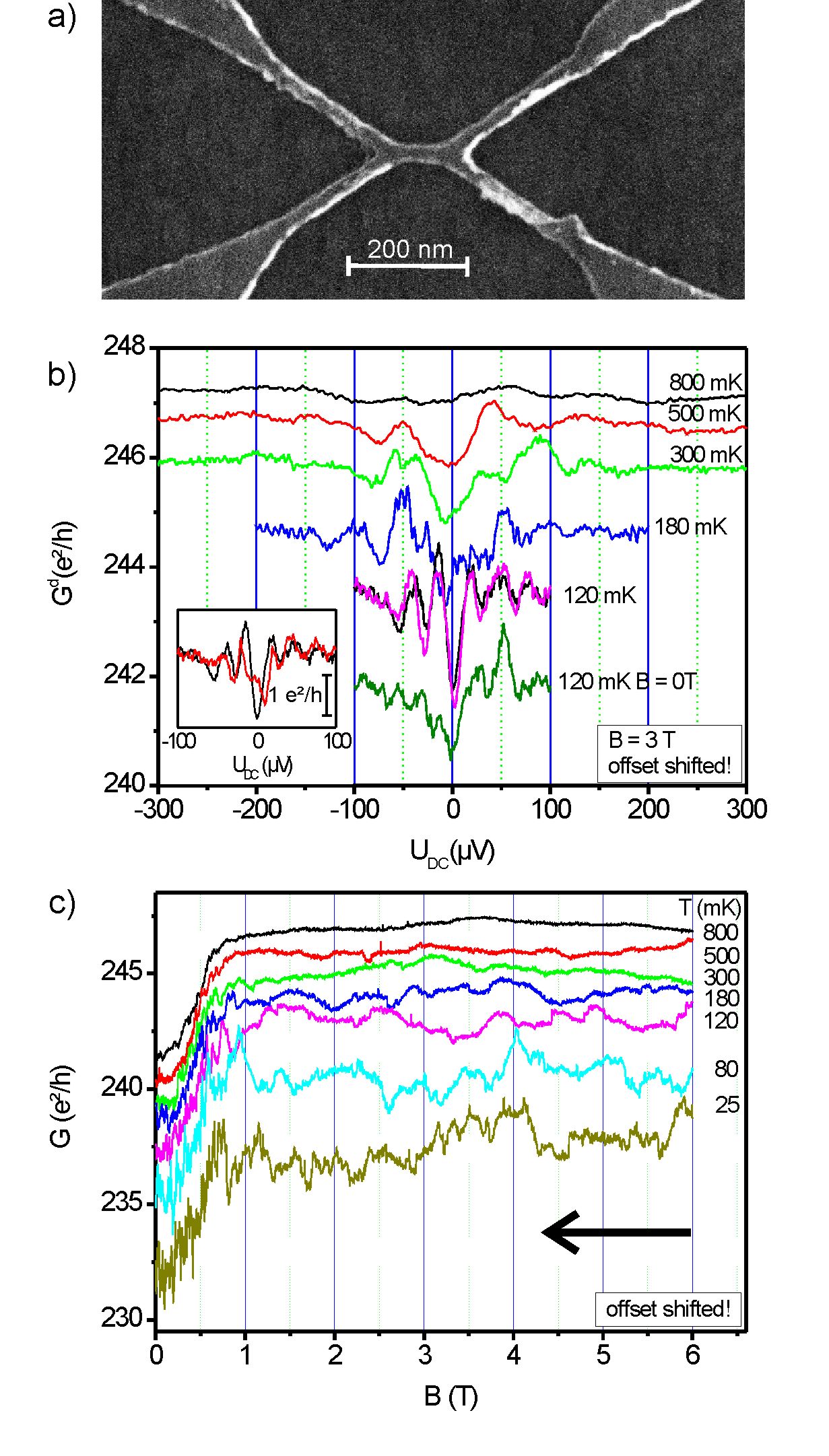}
\caption{a) Electron micrograph of the 145 nm long Permalloy wire.
b) Differential conductance of the 145 nm long Permalloy wire
measured at different temperatures. The applied magnetic field was 3
T (except for the lowest trace: Here $B=0$.). The different traces,
except the 800 mK trace, were shifted for clarity. The
reproducibility is shown for the 120 mK trace (purple and black
trace). The inset shows the differential conductance at 120 mK and 3
T before and after the field was raised to 4 T. The black trace in
the inset corresponds to the black trace at 120 mK in the main
figure. c) Magnetoconductance of the 145 nm long wire in a
perpendicular applied magnetic field at different temperatures. Also
here, the different traces, except the 800 mK trace, were shifted
for clarity. The arrow gives the sweeping direction.}
\end{figure}

The differential conductance $G^d(U_{DC})$ as a function of the
applied bias voltage $U_{DC}$ of the 145 nm long wire is shown in
figure 1b for temperatures $T$ ranging from 120 mK to 800 mK. The
magnetic field applied normal to the sample was 3 T (except for the
lowest trace: Here $B=0$). At all temperatures fluctuations in the
differential conductance are visible, showing that phase coherent
transport takes place in the wire. The reproducibility of the
differential conductance is demonstrated for the $T=120$ mK trace.
With increasing temperature the fluctuation amplitude gets reduced.
This shows that the sample is larger than the phase coherence length
at least at 180 mK. Also with increasing DC-voltage the amplitude of
the fluctuations decreases. This suggests that the applied voltage
causes heating. The correlation voltage $U_C$, the average spacing
between maxima and minima, gives the energy, which is necessary to
change the electron's phase by $\pi$. Here $U_C$ increases with
increasing temperature. The correlation voltage is related to the
dephasing time $\tau_\phi$ by: $|e|U_C=\hbar/\tau_\phi$
\cite{Datta}. Hence the dephasing time decreases with increasing
temperature as expected \cite{Lin}. At $T=120$ mK the correlation
voltage is approx. 17 $\mu$V. This corresponds to a dephasing time
of 40 ps. With the diffusion constant of Permalloy
$D\approx4\cdot10^{-4}$ m$^2$/Vs \cite{Kasai2} we obtain for the
phase coherence length $L_\phi=\sqrt{D\tau_\phi}=130$ nm at 120 mK.
For comparison: The thermal energy $k_BT=10$ $\mu$eV at 120 mK.
Hence thermal broadening does not lead to a suppression of
interference as $k_BT<eU_C$ \cite{Beenakker}.

To investigate the temperature dependency of the dephasing time we
plotted the correlation voltage of the 145 nm long wire in a log-log
diagram versus temperature (see inset of figure 2). The slope of the
correlation voltage can be approximated by 1. This corresponds to a
dephasing time $\tau_\phi\propto 1/T$ and thus to a phase coherence
length $L_\phi\propto 1/\sqrt{T}$. We note that the estimate of the
correlation voltage is associated with a relatively high degree of
uncertainty, as only the region around zero voltage can be used for
an estimate of $U_C$. Hence the slope gives only a rough estimate of
the temperature dependency.

The differential conductance of the 330 nm long wire also exhibits
reproducible fluctuations associated with phase coherent transport
(not shown). At $T=120$ mK the correlation voltage in the 330 nm
long wire is 20 $\mu$V, which is quite close to the correlation
voltage observed in the 145 nm long wire at the same temperature.
Hence the dephasing time and the phase coherence length are quite
similar in both wires, as the measurement of the correlation voltage
gives the dephasing time independent of the sample's geometry. In
the 330 nm long wire the temperature dependence of the correlation
voltage couldn't be estimated properly as the fluctuations disappear
above 180 mK. This is due to damping of the conductance fluctuations
with increasing sample size. The correlation voltage taken at
$T=120$ mK and $T=180$ mK give a temperature dependency in good
agreement with the one obtained in the 145 nm long wire (see inset
of figure 2).

In mesoscopic samples reproducible conductance fluctuations are also
visible in the magnetoconductance $G(B)$, also known as magnetic
fingerprint of the sample \cite{Lee}. In figure 1c the
magnetoconductance of the 145 nm long wire is shown for temperatures
ranging from 25 mK to 800 mK, exhibiting conductance fluctuations
vanishing with increasing temperature. The positive
magnetoconductance at $B=0...0.7$ T is due to the anisotropic
magnetoresistance (AMR) in ferromagnets \cite{McGuire}. For a
magnetization aligned in wire direction (here the easy magnetic
axis) the conductance is smaller than for perpendicular orientation
(here a hard magnetic axis). In contrast to fluctuations of the
differential conductance, the magnetoconductance fluctuations are
not reproducible within several sweeps. As the conductance
fluctuations only appear at low temperatures and decrease with
increasing temperature and increasing wire length, we ascribe them
to phase coherent phenomena. To check whether the fluctuations
originate from time dependent fluctuations, we measured the
conductance at a fixed magnetic field for the time interval of a
magnetic field sweep. Time dependent fluctuations are visible, but
their amplitude is by approx. a factor of 4 lower compared to the
magnetoconductance fluctuations. Additionally, the time scale of
fluctuations is longer in the time-sweep compared to the magnetic
field sweep. Hence we can conclude that the observed conductance
fluctuations, shown in figure 1c, are primarily due to changes of
$B$ but only superimposed by time depending fluctuations. Time
dependent universal conductance fluctuations in Permalloy nanowires
are in the focus of reference \cite{Lee3}. A possible explanation
for irreproducible magnetoconductance traces could be a change of
the scatterer configuration due to magnetostriction. While the
fluctuations in the differential conductance are well reproducible
for a fixed magnetic field (see figure 1b), they get less
reproducible when the magnetic field is changed between the
measurements. This is shown in figure 1b. The two traces in the
inset show the differential conductance at $T=120$ mK and $B=3$ T.
Between the measurements the magnetic field was increased to 4 T and
then reduced to 3 T again (the black trace shows the differential
conductance at the beginning and the red trace after the magnetic
field was varied). This magnetic field change reduces the
reproducibility of the traces compared to the two traces shown in
figure 1b at 120 mK (black and purple), where the magnetic field was
not changed between the measurements. Hence, a change in the
impurity configuration due to magnetostriction is a possible
candidate for the origin of the observed irreproducibility.

\begin{figure}
\includegraphics[width=0.7\linewidth]{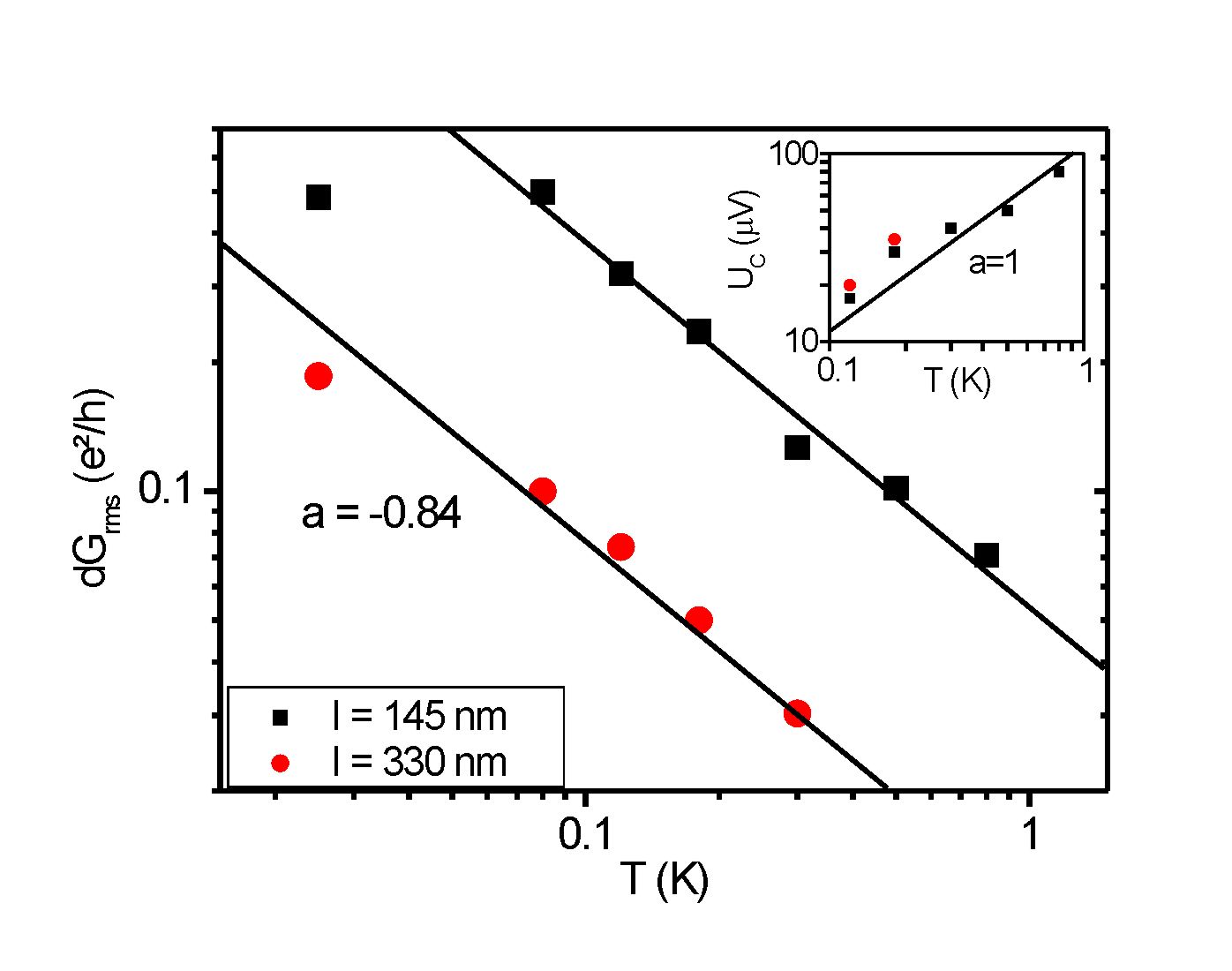}
\caption{Amplitude of conductance fluctuations $dG_{rms}$ measured
in the magnetoconductance of the 145 nm long (black squares) and 330
nm long wire (red circles). The lines are a guide for the eyes and
give a slope of $a=-0.84$. The inset shows the correlation voltage
of both wires for different temperatures. Also here black squares
correspond to the 145 nm long wire and red circles to the 330 nm
long wire. The straight line is a guide for the eyes and gives a
slope of $a=1$.}
\end{figure}

In the low field region ($|B|<0.7$ T), the region where the
magnetization rotates from in plane to perpendicular to the plane,
the conductance fluctuations are more pronounced than in the high
field region. Such a behavior has already been observed in (Ga,Mn)As
wires \cite{Konni,Vila} and ad hoc ascribed to the formation of
domain walls \cite{Vila}. Analyzing the fluctuations in the
differential conductance at zero magnetic field (figure 1b, lowest
trace) one finds, that neither the amplitude nor the correlation
voltage is significantly different compared to $B=3$ T. This implies
that the dephasing time is not much affected due to the presence of
domain walls in the low field region. A similar result was found in
reference \cite{Lee3} analyzing time dependent universal conductance
fluctuations (TDUCF) in Permalloy nanowires. Their analysis of
TDUCFs supports the idea that domain walls act as coherent
scatterers \cite{Lee3}. In the following the analysis is limited to
high magnetic fields ($B=1...6$ T), where the magnetization is
saturated perpendicular to the plane.

The root mean square amplitude of the conductance fluctuations
$dG_{rms}$ of the 145 nm and 330 nm long wires (taken at $B=1...6$
T) is plotted in a log-log diagram versus $T$ in figure 2. Here
$dG_{rms}=\sqrt{\langle (G-\langle G\rangle)^{2}\rangle}$, where
$\langle ...\rangle$ denotes averaging over $B$. For a quasi one
dimensional wire ($w,t<L_\phi<l$) one finds for the amplitude of the
conductance fluctuations \cite{Lee}:
\begin{equation}
dG_{rms}=C\frac{e^2}{h}\left(\frac{L_{\phi}}{l}\right)^{3/2}.
\end{equation}
Here, $l$ is the wire length and $C$ is a constant, with a value
close to or smaller than unity, depending, e.g., on the strength of
spin-orbit coupling \cite{Chandrasekhar} and the applied magnetic
field \cite{Lee}. Equation (1) is applicable to describe the
conductance fluctuations as long as thermal averaging can be ruled
out: $L_\phi<L_T=\sqrt{\hbar D/k_BT}$, which is equivalent to
$eU_C>k_BT$. In our Permalloy wire this is the case as shown above
and equation (1) can be used to extract the phase coherence length.
$dG_{rms}$ of the 145 nm long wire is increasing with decreasing
temperature until it starts to saturate at 80 mK at a value of 0.5
$e^2/h$ (see figure 2). For such a saturation, there are essentially
three possible reasons:

1. The effective electron temperature is higher than the bath
temperature due to a too high measuring current or external
RF-noise.

2. Magnetic impurities lead to a saturation of $L_\phi$ due to
Kondo-scattering as observed in normal metals \cite{Pierre}.

3. The phase coherence length reaches the wire length and the wire
is no longer quasi one dimensional. In that case $dG_{rms}=Ce^2/h$
\cite{Lee}.

As we deal here with a ferromagnet and since saturation is only
observed in the shorter one of the two wires, an intrinsic
saturation of the phase coherence length due to Kondo-scattering, is
unlikely. While the amplitude of the conductance fluctuations of the
25 mK and the 80 mK trace in figure 1c is the same, the correlation
field $B_C$ is smaller at 25 mK. The correlation field defines the
typical field scale on which the conductance fluctuates and is
related to the maximum area enclosed  by a phase coherent
trajectory. In one dimensional samples
$B_C=\frac{h}{e}\frac{1}{wL_\phi}$ \cite{Lee}. We note that the
correlation field might not be a well defined quantity here, as the
impurity configuration might be changed by magnetostriction, as
discussed above. Hence the correlation field can only serve as a
very rough estimate of the phase coherence length. At 80 mK the
correlation field is between $\sim$0.6 T and $\sim$0.8 T. This
corresponds to a phase coherence length of $\sim$150-200 nm. When
extrapolating the correlation voltage $U_C$ (see inset of figure 2)
down to 80 mK, one arrives at a phase coherence length of 160 nm.
This is in very good agreement with the phase coherence length
estimated from the correlation field. Hence, the saturation of
$dG_{rms}$ observed in the 145 nm long wire below 80 mK is most
likely due to an dimensional cross-over from quasi 1-D to 0-D and
the saturation value of $dG_{rms}$ is $\sim0.5e^2/h$.

In both wires the temperature dependency of $dG_{rms}$ can be
approximated by a power law: $dG_{rms}\propto T^{-0.84}$. Using
equation (1), the temperature dependency of the dephasing length is:
$L_\phi\propto T^{-0.55}$. This temperature dependency agrees well
with the one estimated using the correlation voltage $U_C$
($L_\phi\propto T^{-0.5}$). In Permalloy wires noise measurements
reveal a temperature dependency of $L_\phi$ probably steeper than
$T^{-0.5}$ \cite{Lee3}, but the analysis was complicated by several
uncertainties. Also the wires investigated in \cite{Lee3} were quasi
two dimensional, because of the higher temperature ($T>2$ K). Hence
a direct comparison is difficult. In ferromagnetic (Ga,Mn)As wires
the phase coherence length followed a $T^{-0.5}$ dependency
\cite{Konni,Vila} and was associated with critical
electron-electron-scattering (CEEI) \cite{Konni,Review,Dai}. For
Permalloy CEEI seems not to be a suitable candidate for dephasing as
CEEI describes dephasing in a strongly disordered metal near the
metal insulator transition (MIT). In metals far away from the MIT
dephasing is usually ascribed to Nyquist scattering leading to a
phase coherence length $\propto T^{-1/3}$ in 1-D systems
\cite{Altshuler}. So, the microscopic mechanism of dephasing in
Permalloy remains an open issue.

The value of the phase coherence length at 25 mK is approx. 250 nm.
We arrive at this value by analyzing the magnetoconductance
fluctuations of the 330 nm long wire ($L_\phi=180$ nm) or by
extrapolating the magnetoconductance fluctuations of the 145 nm long
wire down to 25 mK ($L_\phi=260$ nm). We note that there is some
uncertainty in determining the length of the wires, especially in
the region of the voltage probes, where the wire widens up. This
uncertainty enters the phase coherence length linearly. Taking the
correlation energy and extrapolating the value of $U_C$ down to 25
mK leads to a phase coherence length of $\sim$260 nm (145 nm long
wire) and $\sim$240 nm (330 nm long wire). This approximation is
independent on the exact wire geometry. In reference \cite{Kasai}
the phase coherence length in Permalloy was extracted from periodic
Aharonov-Bohm oscillations. Their value of $L_\phi=500$ nm taken at
30 mK, is by a factor of 2 larger than the value estimated here. The
difference might be explained by the different material used. The
saturation field of the magnetization observed in a perpendicular
external field, defined by the saturation field of the AMR is in
reference \cite{Kasai} $\sim1.2$ T. For Permalloy
(Ni$_{81}$Fe$_{19}$) the saturation magnetization is 1.0 T. Hence
the saturation field is 1.0 T in an extended film and 0.5 T in a
cigar shaped ellipsoid with infinite aspect ratio (comparable to a
long wire with a very small cross-section) and with the long axis
perpendicular to the external field \cite{Hubert}. For the geometry
used in reference \cite{Kasai} (a 2.5 $\mu$m long wire with a
cross-section of $40\times20$ nm$^2$, having a ring of 500 nm
diameter and the same cross-section in the middle; see reference
\cite{Kasai} for details) one would expect a saturation field above,
but still close to 0.5 T. In our samples the saturation field is
$\sim0.7$ T in the 145 nm and 330 nm long wires (see figure 1c),
$\sim1.0$ T in the extended film (see inset of figure 3a) and
$\sim0.6$ T in the wire array (see inset of figure 3b). These values
are consistent with the ones expected for Permalloy. Hence the
material used in reference \cite{Kasai} seems to be quite different
from the material used here and the values of the phase coherence
length can hardly be compared.

\section{Electron-electron interaction in 1D and 2D samples}

\begin{figure}
\includegraphics[width=0.6\linewidth]{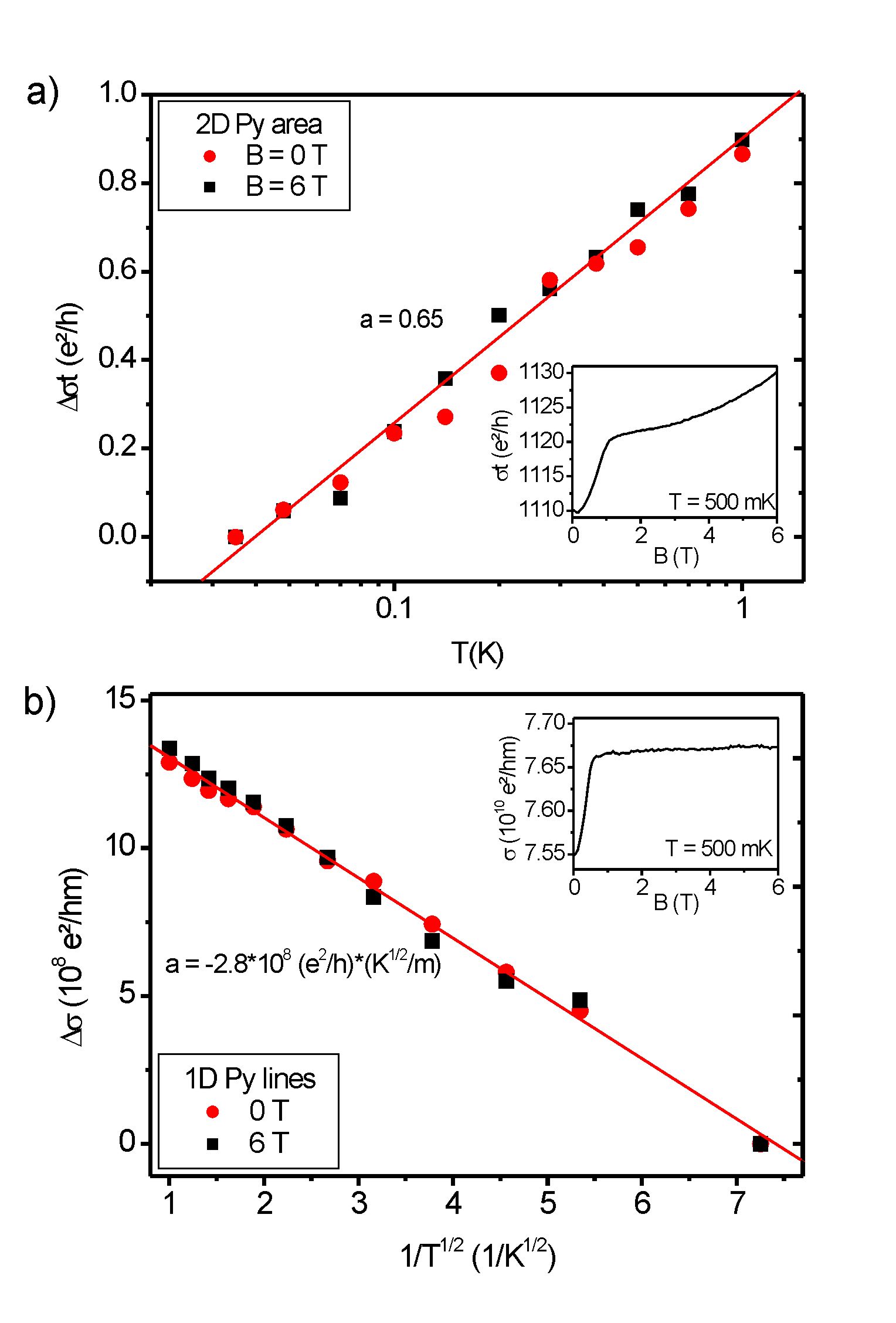}
\caption{a) Change in conductivity $\Delta\sigma$ multiplied with
the layer thickness $t$ of the two-dimensional square at zero field
(red circles) and $B=6$ T (black squares) plotted versus log$T$. The
line is the best linear fit of the 6 T data and has a slope of 0.65.
The inset shows the magnetoconductance at $T=500$ mK. b)
Conductivity change of a quasi one dimensional line array at zero
field (red circles) and $B=6$ T (black squares) plotted versus
$1/\sqrt{T}$. The line is the best linear fit of the 6 T data and
has a slope of $-2.8\cdot10^{8}$
(e$^2$/h)$\cdot(\sqrt{\mathrm{K}}/$m). The inset shows the
magnetoconductance at $T=500$ mK.}
\end{figure}

At low temperatures quantum interference effects like weak
localization (WL) \cite{Bergmann} or electron-electron interaction
(EEI) \cite{Lee2} lead to a reduction of the conductance even in
macroscopic samples. The effect of weak localization originates from
the constructive interference of partial waves traveling on time
reversed paths. This leads to an increased probability of
backscattering and hence reduces the conductance. As a magnetic
field removes time-reversal symmetry, weak localization gets
suppressed in the presence of a magnetic field and the corresponding
magnetoconductance is positive \cite{Bergmann}. In Permalloy wires
and films a positive magnetoconductance in a perpendicular magnetic
field, as expected for WL, is already present due to the anisotropic
magnetoresistance (AMR) \cite{McGuire}. But in contrast to WL the
AMR is independent of the temperature, at least below 1 K, and thus
does not lead to a change in conductivity with decreasing
temperature. Electron-electron interaction, on the other hand,
arises from a modified screening of the Coulomb potential due to the
diffusive propagation \cite{Lee2}. The size and the temperature
dependency of the conductivity correction are very similar for EEI
and WL at zero field \cite{Lee2}, but in contrast to WL, EEI is not
suppressed by a magnetic field. Hence the different effects (AMR, WL
and EEI) can be distinguished quite effectively by comparing the
conductivity decrease at different magnetic fields. In case of WL
the conductivity at zero field is expected to decrease more
precipitously with decreasing $T$ compared to the case with finite
$B$.

As ferromagnets have an internal magnetic induction, the question
arises whether WL can be observed in a ferromagnet at all. Up to now
several experimental works explore this question, showing that WL is
absent in Co \cite{Brands,Brands2}, Fe \cite{Brands2}, Ni \cite{Ono}
and Co/Pt multilayers \cite{Brands3}. In the ferromagnetic
semiconductor (Ga,Mn)As, having a rather small internal magnetic
induction, weak localization was observed \cite{WL,Rozkinson}. As
theory claims that WL gets not suppressed by internal magnetic
induction \cite{Dugaev,Sil}, the question arises why WL can be
observed in (Ga,Mn)As but not in Co, Fe, Ni, or Co/Pt multilayers.
One important difference between these materials is the phase
coherence length. While $L_\phi\approx160$ nm at 20 mK is quite
large in (Ga,Mn)As \cite{Review}, the phase coherence length is
comparably small in Co ($L_\phi\approx30$ nm at 30 mK \cite{Wei})
and Ni ($L_\phi\approx80$ nm at 30 mK \cite{Kasai2}). Hence
Permalloy is a quite promising candidate for exploring WL as it
combines a strong internal magnetic induction ($H_{Int}=1.0$ T
compared to 40 mT in (Ga,Mn)As \cite{Ohno}) with a relatively large
phase coherence length of approx. 250 nm at 25 mK.

To study Permalloy for WL/EEI experiments, we fabricated an extended
square ($w=l=4$ $\mu$m, $t=$15 nm) and an array of 6 wires in
parallel ($w=20$ nm, $l=4$ $\mu$m, $t=15$ nm) and measured their
conductivity at different temperatures with and without an applied
perpendicular magnetic field. The square is expected to behave quasi
two-dimensional, as $w,l>L_\phi,L_T>t$ and the wire array is
expected to behave quasi one-dimensional, as $l>L_\phi,L_T>w,t$ at
temperature below 1 K. The magnetoconductance of the square and the
wire array are shown in the insets of figure 3 for $T=500$ mK. Due
to ensemble averaging universal conductance fluctuations are
suppressed in both systems. For both samples the AMR is visible up
to 1 T (area) and 0.6 T (wires) as expected for Permalloy and the
corresponding saturation fields \cite{Hubert}. The temperature
dependency of the conductivity change of the area, relative to
$T=35$ mK, is shown in figure 3a for $B=6$ T (black squares) and
zero field (red circles). At both magnetic fields the conductivity
follows a logarithmical temperature dependency with a slope
independent of the applied magnetic field. Hence WL can be ruled out
as origin of the conductivity decrease, as WL gets suppressed by an
external field. EEI interaction leads to a conductivity decrease
independent of the applied magnetic field. For the conductivity
decrease due to EEI one obtains in two-dimensional films
\cite{Lee2}:
\begin{equation}
\Delta\sigma
t=\frac{F^{2D}}{\pi}\frac{e^2}{h}\mathrm{log}\frac{T}{T_0},
\end{equation}
with a screening factor $F^{2D}$. In our Permalloy square one
obtains a screening factor $F^{2D}=2.0$, using a slope of 0.65 as
shown in figure 3a. This value of $F^{2D}$ is in excellent agreement
with the screening factors found in other ferromagnets: For Co
$F^{2D}=2.0...2.6$ \cite{Brands,Brands2}, for (Ga,Mn)As
$F^{2D}=1.8...2.6$ \cite{EEI} and for Co/Pt multilayers $F^{2D}=2.5$
\cite{Brands3}. In Fe and Ni the screening factors are comparable to
the screening factor observed in Co \cite{Brands2}. We note that the
effect of EEI is rather small in the quasi two-dimensional film. The
relative conductivity change $\Delta\sigma/\sigma$ from 1 K to 35 mK
is only $\sim8\cdot 10^{-4}$.

The magnetoconductance of the wire array, taken at 500 mK, is shown
in the inset of figure 3b. Also in the wire array the AMR is visible
but saturates already at $\approx0.6$ T. This is due to the
different shape anisotropy compared to an extended film
\cite{Hubert}. The conductivity decrease with respect to 20 mK is
plotted in figure 3b for $B=6$ T and zero field. Also in the wire
array the conductivity decrease is independent of the applied
magnetic field. Hence WL is absent also in the wire array. The
temperature dependency of the conductivity decrease is rather
different in the wire array and the Py-square. In the wire array
$\Delta\sigma$ follows a $1/\sqrt{T}$ dependency as it is expected
for EEI in quasi one-dimensional samples \cite{Lee2}:
\begin{equation}
\Delta\sigma =-\frac{F^{1D}}{\pi
A}\frac{e^2}{\hbar}\sqrt{\frac{\hbar D}{k_BT}},
\end{equation}
with a screening factor $F^{1D}$, the wire cross-section A and the
thermal diffusion length $L_T=\sqrt{\frac{\hbar D}{k_BT}}$. Using
$D=4\cdot 10^{-4}$ m$^2$/Vs \cite{Kasai2} and
$\delta\sigma\sqrt{T}=-2.8\cdot10^8$
(e$^2$/h)$\cdot$($\sqrt{\mathrm{K}}$/m) (see figure 3b) one obtains
as screening factor: $F^{1D}=0.77$. Also the screening factor
$F^{1D}$ is in excellent agreement compared to the screening factors
obtained for other quasi 1-D ferromagnets ($F^{1D}=0.83$ in Ni
\cite{Ono} and $F^{1D}=0.72...0.80$ in (Ga,Mn)As \cite{EEI}). In the
quasi one-dimensional wire array the effect of EEI becomes quite
large compared to the quasi two-dimensional film, as predicted by
theory \cite{Lee2}. In 1D the conductivity decreases by approx. 1.1
\% from 1 K to 35 mK. This is by a factor of 14 larger than the
conductivity decrease in the quasi two-dimensional film.

\section{Summary}

We investigated quantum interference effects in mesoscopic Permalloy
films and wires at milikelvin temperatures. Our analysis of
universal conductance fluctuations in single wires reveals a phase
coherence length of $\sim250$ nm at 25 mK. Compared to other
ferromagnets this value is relatively large (In Co $L_\phi=30$ nm at
30 mK \cite{Wei}, in Ni $L_\phi=80$ nm at 30 mK \cite{Kasai2}). A
possible explanation was given by Kasai \textit{et al.}
\cite{Kasai2}: In ferromagnets the phase coherence length gets
probably reduced with increasing magnetocrystalline anisotropy
energy. The observed temperature dependency of the dephasing time
$\propto T^{-1/2}$ is stronger than expected for Nyquist scattering
in 1-D systems \cite{Altshuler} ($\propto T^{-1/3}$). A similar
result was already obtained by Lee \textit{et al.} \cite{Lee3}
investigating TDUCF in Permalloy. A microscopic mechanism for the
$T^{-1/2}$ temperature dependence is still missing. Although the
phase coherence length is relatively large, the conductance in 1-D
and 2-D Permalloy is not affected by weak localization. The
decreasing conductance with decreasing temperature can be well
described by EEI, as already shown for other ferromagnetic metals
\cite{Brands,Brands2,Brands3,Ono}. The excellent agreement of the
screening factors found in the ferromagnets investigated up to now,
underline the universal character of EEI. The size of the WL
correction is expected to be of the same order as EEI at zero field
\cite{Lee2}. Hence our data show that WL is strongly suppressed or
even absence in Permalloy. The absence of a Cooperon contribution in
Permalloy, the process leading to WL, has already been proposed by
Lee \textit{et al.} investigating TDUCF \cite{Lee3}. Although theory
predicts the existence of WL in ferromagnets \cite{Dugaev,Sil}, it
seems that the Cooperon contribution is suppressed in ferromagnets,
independent on the value of the phase coherence length. The process
which could lead to such a suppression is still unknown. Hence,
further investigations, experimental as well as theoretical, are
necessary to clarify the role of the internal magnetic induction on
the phase coherent transport in general, and on the Cooperon in
particular.

Acknowledgement: This work was financially supported by the Deutsche
Forschungsgemeinschaft (DFG) via SFB 689.



\end{document}